\documentclass[final,5p,times,twocolumn]{elsarticle}
\usepackage{amsmath,amssymb,amsfonts}
\usepackage{graphicx}
\usepackage{hyperref}
\usepackage{multirow}
\usepackage{mciteplus}
\mciteErrorOnUnknownfalse

\hypersetup{
  hidelinks,
  pdftitle={Coupled-channel Omnes matrix for the D-wave isoscalar pi-pi/K-Kbar system and its application to J/psi radiative decays},
  pdfauthor={Igor Danilkin, Oleksandra Deineka, Emilie Passemar, Marc Vanderhaeghen}
}

\makeatletter
\def\bibinfo@X@title#1,{\ignorespaces}
\def\ps@pprintTitle{%
  \let\@oddhead\@empty
  \let\@evenhead\@empty
  \def\@oddfoot{\hfil\thepage\hfil}%
  \let\@evenfoot\@oddfoot}
\makeatother

\begin{document}

\title{Coupled-channel Omn\`es matrix for the $D$-wave isoscalar $\pi\pi/K\bar K$ system and its application to $J/\psi\to\pi^{0}\pi^{0}\gamma,\,K_{S}K_{S}\gamma$
}

\author[Mainz]{Igor Danilkin}
\author[Mainz]{Oleksandra Deineka}
\author[Indiana,Valencia]{Emilie Passemar}
\author[Mainz]{Marc Vanderhaeghen}

\address[Mainz]{Institut f\"ur Kernphysik and 
PRISMA$^+$ Cluster of Excellence, Johannes Gutenberg Universit\"at, 55099 Mainz, Germany}
\address[Indiana]{\it Physics Department, Indiana University, Bloomington, Indiana 47405, USA}
\address[Valencia]{\it Departament de F\'{i}sica Te\`{o}rica, Instituto de F\'{i}sica Corpuscular,

Universitat de Val\`encia -- Consejo Superior de Investigaciones Cient\'{i}ficas,

Parc Cient\'{i}fic, Catedr\'{a}tico Jos\'{e} Beltr\'{a}n 2, E-46980 Paterna, Valencia, Spain}

\date{}

\begin{abstract}

In this work, we construct the $D$-wave isoscalar $\pi\pi/K\bar K$ coupled-channel Omn\`es matrix, formulated to satisfy unitarity, analyticity, and the appropriate asymptotic behavior. We employ a two-channel $K$-matrix model containing poles associated with the $f_{2}(1270)$ and $f_{2}'(1525)$ resonances. The resulting unitary scattering matrix, which reproduces the experimental $\pi\pi\to\pi\pi$ and $\pi\pi\to K\bar K$ data and PDG information, serves as input to the homogeneous two-channel Muskhelishvili-Omn\`es equation. We compare our Omn\`es matrix with previous constructions based on $\pi\pi\to K\bar K$ phases extracted from sums of Breit-Wigner amplitudes. The Omn\`es matrix developed here provides a reliable dispersive input for form-factor calculations and resonance studies in the tensor-meson sector. As an application, we show that it enables a simultaneous and accurate description of the BESIII $J/\psi\to\pi^{0}\pi^{0}\gamma$ and $J/\psi\to K_{S}K_{S}\gamma$ spectra in the $J=2$ electric-dipole (E1) partial wave.
\end{abstract}
	
\maketitle

\section{Introduction}\label{sec:intro}

Hadronic final-state interactions play a crucial role in determining resonance properties and in dispersive analyses of form factors. The Omn\`es function (or matrix, in coupled channels) is a central tool to incorporate these interactions in a way consistent with unitarity and analyticity. In this paper we focus on the $D$-wave isoscalar $\pi\pi/K\bar K$ system, which is dominated by the well-known $f_2(1270)$ resonance in the $\pi\pi$ channel and by $f'_2(1525)$ in the $K\bar K$ channel.

Most dispersive applications so far have relied on a single-channel Omn\`es function in the tensor isoscalar channel (see e.g. \cite{Garcia-Martin:2010kyn,Hoferichter:2015hva,Danilkin:2018qfn,Hoferichter:2019nlq,Danilkin:2019opj,Lellmann:2025aje}). A genuinely coupled-channel treatment of the $\pi\pi/K\bar K$ system is required to describe consistently the dynamics of the $f_2(1270)$-$f'_2(1525)$ region. Several recent studies have required a $D$-wave coupled-channel $\pi\pi/K\bar K$ Omn\`es function as input. Tarr{\'u}s Castell{\`a} \emph{et al.} \cite{TarrusCastella:2021pld} constructed an Omn\`es solution for the $IJ=02$ channel to analyze transitions between exotic and standard bottomonia. Several years later, Cao \emph{et al.} \cite{Cao:2025dkv} employed a coupled-channel Omn\`es representation in a dispersive study of the pion and nucleon gravitational form factors. Despite these applications, the construction of a reliable D-wave $\pi\pi/K\bar K$ Omn\`es matrix remains challenging. Direct scattering data in the $K\bar{K}$ channel do not exist, and the available information is largely limited to branching fractions and resonance parameters summarized by the PDG~\cite{ParticleDataGroup:2024cfk}. Nevertheless, lattice QCD studies at unphysical pion masses \cite{Briceno:2017qmb} have shown that the D-wave $\pi\pi$ and $K\bar{K}$ phase shifts can be successfully described using a $K$-matrix parametrization with two poles and a tiny background contribution.

Motivated by these considerations, in this paper we construct a D-wave isoscalar $\pi\pi/K\bar K$ coupled-channel Omn\`es matrix based on a unitary two-channel $K$-matrix model incorporating the $f_2(1270)$ and $f'_2(1525)$ resonances. Fitting this model to experimental $\pi\pi \to \pi\pi$ and $\pi\pi \to K\bar{K}$ data and PDG information yields a unitary scattering $t$-matrix that serves as input for solving the homogeneous two-channel Muskhelishvili-Omn\`es problem. This construction addresses several shortcomings of earlier approaches \cite{TarrusCastella:2021pld,Cao:2025dkv}, which relied on $\pi\pi \to K\bar K$ phases obtained from dispersively improved sums of Breit-Wigner amplitudes \cite{Pelaez:2018qny,Pelaez:2020gnd}. While the CFD step incorporates some dispersive constraints (analyticity/crossing), it is not obtained from an explicit unitary $2\times2$ coupled-channel fit and therefore does not enforce coupled-channel unitarity by construction; in particular, the phase of the $D$-wave transition amplitude remains model-dependent. Furthermore, we will show that the $K\bar{K}$ phase shift implied by those phases is physically unreasonable: the $K\bar{K}$ phase shift is small and even negative, whereas from PDG one expects a sizable phase shift due to the $f'_2(1525)$ resonance.

As an illustration of the utility of the resulting Omn\`es matrix, we apply it to the radiative decays $J/\psi\to\pi^{0}\pi^{0}\gamma$ and $J/\psi\to K_{S}K_{S}\gamma$. We demonstrate that our dispersive representation of the $J = 2$ electric-dipole (E1) amplitudes leads to a simultaneous and accurate description of both decay spectra.

\section{Formalism}
\label{sec:Formalism}
\subsection{Scattering amplitude and unitarity}
In two-body scattering, the starting point is the partial-wave (p.w.) expansion of the invariant amplitude between channels $a$ and $b$,
\begin{equation}\label{eq:pw-expansion}
T_{ab}(s,t)
=16\pi\,\mathcal N_{ab}
\sum_{J=0}^{\infty} (2J+1)\,(p_a p_b)^J\, t^{(J)}_{ab}(s)\,P_J(\cos\theta),
\end{equation}
where $s$ and $t$ are the Mandelstam variables, $\theta$ is the centre-of-mass scattering angle, $P_J$ are Legendre polynomials, and $p_a$ denotes the centre-of-mass (c.m.) three-momentum in channel $a$. In what follows we fix a given $J$ and drop the superscript. For the isoscalar D-wave system we consider two coupled channels $a,b=1,2$ corresponding to $1\equiv\pi\pi$, $2\equiv K\bar K$. The normalization constants are chosen as $ 
\mathcal{N}_{11}=2$, $\mathcal N_{12}=\sqrt{2}$, $\mathcal N_{22}=1$ \cite{Garcia-Martin:2010kyn}, such that the partial-wave amplitudes satisfy a simple unitarity relation for channels with identical and non-identical particles
\begin{align}
  &\text{Im}\,\boldsymbol{t}^{-1}(s) = -\boldsymbol{\rho}(s)\,, \nonumber \\
  &\boldsymbol{\rho}(s)=
  \begin{pmatrix}
     \rho_1(s)\,\theta(s-s_{\mathrm{th}1}) & 0 \\
     0 & \rho_2(s)\,\theta(s-s_{\mathrm{th}2})
   \end{pmatrix},
  \label{eq:Imtinv}
\end{align}
where $s_{\mathrm{th}a}$ denotes the threshold of channel $a$. The corresponding p.w. $S$-matrix is
\begin{equation}
 \boldsymbol{S}(s) = \boldsymbol{1} + 2i\,\sqrt{\boldsymbol{\rho}(s)}\,\boldsymbol{t}(s)\sqrt{\boldsymbol{\rho}(s)}\,.
  \label{eq:Sfromt}
\end{equation}
In Eq.~\eqref{eq:pw-expansion} the threshold factors $(p_a\,p_b)^J$ are pulled out explicitly, so that the partial-wave amplitudes have good analytic properties. In this normalization the phase-space factors entering unitarity take the form
\begin{equation}\label{eq:sigma}
\rho_a(s)=(p_a(s))^{2J}\,\sigma_a(s),
\qquad 
\sigma_a(s)\equiv \frac{2p_a(s)}{\sqrt{s}}\,.
\end{equation}

When both channels are open, unitarity and time-reversal invariance imply that $\boldsymbol{t}(s)$ is symmetric and can be parametrized in terms of three real functions of $s$. A convenient choice is the phase shifts $\delta_1(s)$ and $\delta_2(s)$ in the two channels, together with the modulus of the transition amplitude $|t_{12}(s)|$. In terms of these quantities, the most general solution of the two-open-channel unitarity constraints can be written as
\begin{equation}
  \boldsymbol{t}(s) =
  \begin{pmatrix}
    \dfrac{\eta(s)\,e^{2i\delta_1(s)}-1}{2i\,\rho_1(s)} &
    |t_{12}(s)|\,e^{i\delta_{12}(s)} \\[1.0em]
    |t_{12}(s)|\,e^{i\delta_{12}(s)} &
    \dfrac{\eta(s)\,e^{2i\delta_2(s)}-1}{2i\,\rho_2(s)}
  \end{pmatrix}\,,
  \label{eq:t-matrix-param}
\end{equation}
where $\eta(s)$ is the inelasticity parameter. Above the second
threshold, the unitarity relation
fixes $\eta(s)$ in terms of the transition probability as
\begin{equation}
  \eta(s)
  = \sqrt{1-4\,\rho_1(s)\,\rho_2(s)\,|t_{12}(s)|^2\,
          \theta(s-s_{\mathrm{th}2})}\,.
  \label{eq:eta-from-t12}
\end{equation}
Likewise, the phase of the transition amplitude satisfies
\begin{equation}
  \delta_{12}(s) = \delta_1(s) + \delta_2(s)\,,
\end{equation}
reflecting the standard two-channel parametrization.
In the elastic region, $s<s_{\mathrm{th}2}$, this reduces to
$\delta_{12}(s) = \delta_1(s)$, in accordance with Watson's theorem.

\subsection{Two-channel Muskhelishvili-Omn\`es problem}
\label{sec:MO}
The Omn\`es matrix $\boldsymbol{\Omega}(s)$ is defined as the solution of a homogeneous $2\times2$ Muskhelishvili-Omn\`es (MO) problem that encapsulates the right-hand cut contributions of the scattering $t$-matrix
\begin{align}
\label{eq:MO-2ch}
&\boldsymbol{\Omega}(s)
  = \int_{s_{\mathrm{th}1}}^{\infty}
      \frac{ds'}{\pi}\frac{\mathrm{Im}\,\boldsymbol{\Omega}(s')}{s'-s}\,,\nonumber \\
&\mathrm{Im}\,\boldsymbol{\Omega}(s)
   = \boldsymbol{t}^{\!*}(s)\,\boldsymbol{\rho}(s)\,\boldsymbol{\Omega}(s)    \,.  
\end{align}
An unsubtracted dispersion relation implies the asymptotic behavior $\Omega(s)\sim s^{-1}$, which is often required in form-factor applications \cite{Lepage:1980fj} (for applications see e.g. \cite{Moussallam:2007qc,Celis:2013xja,Albaladejo:2015aca,Yao:2018tqn}). In the one-channel case, Eq. \eqref{eq:MO-2ch} can be solved in closed form and reduces to the well-known expression
\begin{equation}
\Omega(s)=\exp\left(\frac{s}{\pi}\int_{s_{\mathrm{th}1}}^{\infty} \frac{ds'}{s'}\frac{\delta(s')}{s'-s}\right)\,,
\end{equation}
where $\delta(s)$ is the elastic phase shift.
In the coupled-channel case, no closed analytic expression is known for the individual entries of $\boldsymbol{\Omega}(s)$ and one must solve Eq. (\ref{eq:MO-2ch}) numerically. For practical computations it is convenient to work with the real part of $\boldsymbol{\Omega}(s)$. Using standard techniques, the MO system can be rewritten as
\begin{align}
 \label{eq:ReOmega-2ch}
&\mathrm{Re}\,\boldsymbol{\Omega}(s)
  = \mathrm{pv}\!\int_{s_{\mathrm{th}1}}^{\infty}
      \frac{ds'}{\pi}\,
      \frac{\boldsymbol{X}(s')\,\mathrm{Re}\,\boldsymbol{\Omega}(s')}{s'-s}\,\nonumber \\
 &     \mathrm{Im}\,\boldsymbol{\Omega}(s)\equiv \boldsymbol{X}(s)\,\mathrm{Re}\,\boldsymbol{\Omega}(s)\,,
\end{align}
where ``pv'' denotes the Cauchy principal value and\footnote{The simplified form 
$\boldsymbol{X}(s)=\text{Im}\,\boldsymbol{t}(s)\,[\text{Re}\,\boldsymbol{t}(s)]^{-1}$ used in \cite{Yao:2018tqn} is only valid when all channels are open.}
\begin{align}
\label{eq:X-def}
\boldsymbol{X}(s) &= \big[\boldsymbol{1} - i\,\boldsymbol{t}^*(s)\,\boldsymbol{\rho}(s)\big]^{-1} \boldsymbol{t}^*(s)\,\boldsymbol{\rho}(s)\nonumber \\
&=  i\,\left[\boldsymbol{1}-\bigl(\boldsymbol{1}-i\,\boldsymbol{t}^{\!*}(s)\,\boldsymbol{\rho}(s)\bigr)^{-1}\right]\,.
\end{align}  
Since the MO kernel depends only on the product $\boldsymbol{t}^*(s)\,\boldsymbol{\rho}(s)$, in the single-channel case moving D-wave threshold factors between the partial-wave amplitude and the phase space does not change $\Omega(s)$. However, this is no longer true in the coupled-channel case, where the off-diagonal elements of the MO kernel are affected. For $s>s_{\mathrm{th}2}$ one finds
\begin{align}\nonumber
\boldsymbol{t}^*(s)\,\boldsymbol{\rho}(s)\overset{s>s_{\mathrm{th}2}}{=}\begin{pmatrix}
\dfrac{1-\eta\,e^{-2i\delta_{1}}}{2i} & \frac{\sqrt{1-\eta^2}}{2}\,\sqrt{\frac{\rho_2}{\rho_1}}\, e^{-i\delta_{12}} \\
\frac{\sqrt{1-\eta^2}}{2}\,\sqrt{\frac{\rho_1}{\rho_2}} \,e^{-i\delta_{12}} & \dfrac{1-\eta\,e^{-2i\delta_{2}}}{2i} 
\end{pmatrix}\,,
\end{align}
which explicitly depends on the ratios $\sqrt{\rho_2/\rho_1}$ and $\sqrt{\rho_1/\rho_2}$.
Therefore, it is essential to work with partial waves that are free of kinematical constraints, as done in \cite{Cao:2025dkv} (in contrast to \cite{TarrusCastella:2021pld}).

The real-part equation (\ref{eq:ReOmega-2ch}) is discretized following the procedure described in Ref.~\cite{Moussallam:1999aq} and then solved with the help of a singular-value decomposition (for details see also Refs.~\cite{TarrusCastella:2021pld,Celis:2013xja}). To ensure the existence of a unique MO matrix with $\boldsymbol{\Omega}(s)\sim s^{-1}$, in addition to the boundary condition $\boldsymbol{\Omega}(0)=\boldsymbol{1}$, one must impose \cite{Muskhelishvili-book, Moussallam:1999aq}
\begin{equation}\label{eq:constraints}
\sum_{a=1}^{n_{\text{ch}}}\delta_a(\infty) = n_{\text{ch}}\pi,
\qquad
|t_{ab}(\infty)|=0 \ (a\neq b)\,,
\end{equation}
where $n_{\text{ch}}=2$ is the number of considered channels. Note that $|t_{12}(s\to \infty)|$ should drop sufficiently fast so that $\eta(\infty)=1$. Although the full matrix solution of Eq.(\ref{eq:ReOmega-2ch}) is obtained numerically, its determinant can be written in closed form as
\begin{equation}
\det\boldsymbol{\Omega}(s)
= \exp\left(\frac{s}{\pi}\int_{s_{\text{th}1}}^{\infty}ds'
\frac{\delta_{12}(s')}{s'(s'-s)}\right),
\end{equation}
which provides a stringent internal numerical check. In the following analyses, this check is satisfied at the sub-percent level.

Once $\boldsymbol{\Omega}(s)$ is known, any physical form factor or an amplitude $\boldsymbol{F}(s)$ with the same rescattering can be written as $\boldsymbol{F}(s)= \boldsymbol{\Omega}(s)\,\boldsymbol{P}(s)$, where $\boldsymbol{P}(s)$ is a $(2\times 1)$ vector that contains polynomial terms and/or left-hand-cut contributions fixed by the specific form-factor or production process under consideration.

\subsection{Limitations of coupled-channel Omn\`es solution}

The ideal strategy for solving the coupled Muskhelishvili-Omn\`es equations (\ref{eq:ReOmega-2ch}) is to use experimental data as input. However, in practice, there is often no experimental information covering all required quantities, or existing measurements are mutually inconsistent. Moreover, as can be seen from Eqs. (\ref{eq:MO-2ch}) and (\ref{eq:ReOmega-2ch}), the dispersion integral goes over the unitarity cut starting from the lowest threshold. Therefore, in a two-channel problem one needs to know $t_{12}(s)$ outside the physical scattering region. Watson's theorem fixes the phase below the inelastic threshold, but the modulus $|t_{12}(s)|$ is not observable, so one has to rely on a model parametrization.

In the applications to the tensor $IJ=02$ $\pi\pi/K\bar{K}$ system in Refs.~\cite{TarrusCastella:2021pld,Cao:2025dkv}, the hadronic input is taken from \cite{Garcia-Martin:2011iqs} or \cite{Bydzovsky:2016vdx} for the $\pi\pi$ phase shift $\delta_1(s)$, while the modulus $|t_{12}(s)|$ and its phase $\delta_{12}(s)$ are taken from the dispersively constrained Breit-Wigner-like fit to $\pi\pi \to K\bar{K}$ data \cite{Pelaez:2018qny} or its updated version \cite{Pelaez:2020gnd}. These inputs are used up to $2~\text{GeV}$ and then continued to infinity, as required by Eq.~(\ref{eq:constraints}). While this sounds straightforward, several important issues arise when one attempts to use such an approach. These problems are already visible in the better studied scalar $IJ=00$ system and become more severe in the tensor channel. Below we summarize the main limitations, which motivate our current treatment.

\begin{enumerate}
\item \textbf{Input consistency:}  
In practice, $\delta_1$ and $t_{12}$ are often taken from different analyses. A classic example is the $IJ=00$ system \cite{Daub:2012mu,Hoferichter:2012wf}, where the input for $|t_{12}|$ from the Roy-Steiner analysis of \cite{Buettiker:2003pp} was readjusted to match the $\pi\pi$ phase shift from the Bern Roy-equation solution \cite{Caprini:2011ky} in order to maintain consistency. In contrast, for the $IJ=02$ system no MO problem with a finite matching point was solved in Refs.~\cite{Pelaez:2018qny,Pelaez:2020gnd} to obtain $t_{12}(s)$ in the unphysical region. Consequently, both implementations \cite{TarrusCastella:2021pld,Cao:2025dkv} rely on the so-called CFD parametrizations \cite{Pelaez:2018qny,Pelaez:2020gnd} of $t_{12}$ for all $s$. Treating $\delta_1$ and $t_{12}$ as if they were fully consistent input to a two-channel Omn\`es problem introduces an uncontrolled systematic uncertainty, especially in the unphysical region below the $K\bar{K}$ threshold where no data are available. A simplification compared to the $IJ=00$ case is that in $IJ=02$ there are no resonances below the $K\bar{K}$ threshold: there is only the tail of the $f_2(1270)$ contribution.\\

\item \textbf{Analytic continuation and unitarity:}  
Even if $\delta_1$ and $|t_{12}|$ are mutually consistent on the real axis, this does not guarantee that the poles extracted from different channels coincide in the complex plane. This was, for instance, explicitly observed in the scalar sector \cite{Pelaez:2022qby}: analytic continuation of the $\pi\pi \to K\bar{K}$ amplitude led to a pole position of $f_0(1370)$ that differs somewhat from that obtained from $\pi\pi \to \pi\pi$. Unless one imposes coupled-channel unitarity at the level of fitting to scattering data, this problem will remain, because the different channels are fitted to different experimental data sets, which need not be perfectly compatible. Without a dynamical framework that enforces coupled-channel unitarity, such as a $K$-matrix or $N/D$ approach, the Omn\`es matrix ``inherits'' these inconsistencies. In the tensor channel, where only the modulus of the D-wave has been measured experimentally and the phase is completely model-dependent, this issue is particularly important. \\

\item \textbf{Two-channel unitarity:} 
It is not advisable to use $\delta_1$ and $t_{12}$ from data far above the energy region where a two-channel description remains valid, i.e., where additional inelastic channels become important. In extreme cases, the inelasticity parameter can even become complex, signaling a breakdown of two-channel unitarity in the input. This happens, for instance, in the $IJ=00$ case if the CFD$_\mathrm{C}$ solution of Ref.~\cite{Pelaez:2020gnd} is used for $|t_{12}|$ above the $K\bar{K}$ threshold. \\

\item \textbf{Sum of eigen-phases constraint:}  
Unsubtracted coupled-channel MO solutions require assumptions about the asymptotic behavior of the $S$-matrix eigen-phases, as shown in Eq.(\ref{eq:constraints}). In practice, most analyses \cite{Cao:2025dkv, Albaladejo:2015aca,Yao:2018tqn,Hoferichter:2012wf,Lu:2020qeo} implement this constraint by interpolating between a value $\delta_i(s_0)$ at some matching point $s_0$ and a prescribed limit $\delta_i(\infty)$, usually taken to be an integer multiple of $\pi$ \cite{Moussallam:1999aq}\footnote{A smooth interpolation version of \cite{Moussallam:1999aq} is given in \cite{Gonzalez-Solis:2019iod}}. However, phenomenology rarely provides enough information to determine how the total phase sum should be distributed among the individual eigen-phases. Different choices can lead to significantly different Omn\`es solutions, and this ambiguity is typically not reflected in the quoted uncertainties. The $IJ=02$ $\pi\pi/K\bar{K}$ system provides a clear illustration. The analyses of Refs.~\cite{TarrusCastella:2021pld,Cao:2025dkv} use essentially the same input below $2~\text{GeV}$, but assign different high-energy limits to the eigen-phases: $\delta_1 \to 2\pi$, $\delta_2 \to 0$ \cite{TarrusCastella:2021pld} versus $\delta_1 \to \pi$, $\delta_2 \to \pi$ \cite{Cao:2025dkv}. As a result, the corresponding Omn\`es matrices differ dramatically, even though both are formally consistent with the sum of eigen-phases constraint\footnote{In principle, one might hope that polynomial prefactors in the MO representation could absorb the ambiguity associated with the eigen-phase asymptotics, as suggested in \cite{Yao:2018tqn}. In practice, however, only very low-degree polynomials are used, and their coefficients are constrained by additional information (e.g. chiral low-energy constants or lattice input). Consequently, the freedom to compensate for sizable deformations of the Omn\`es matrix is rather limited.}.

\end{enumerate}

The issues 1-3 can be resolved by using a properly constructed unitary two-channel $t$-matrix model with parameters fitted to experimental data, as was done, for instance, in \cite{Dai:2014zta} for the $IJ=00$ system. The difference between Omn\`es matrices constructed using direct input ($\delta_1$ and $t_{12}$) and those based on the $K$-matrix model is sometimes taken as an estimate of systematic uncertainties \cite{Albaladejo:2016mad, Chen:2019mgp}. Regarding the high-energy constraint (point 4), the limit $\delta_1 \to 2\pi$, $\delta_2 \to 0$ looks phenomenologically plausible for the $IJ=00$ system, and a $K$-matrix model with two poles (corresponding to the $f_0(500)/f_0(980)$) and Chew-Mandelstam phase space satisfies this constraint as well. However, there is a clear limitation to adding an additional channel (e.g., $\eta\eta$), which would force $\sum \delta_a \to 3\pi$. To avoid this, one can relax $\Omega(s)\sim s^{-1}$ and instead solve a once-subtracted version of Eq.~(\ref{eq:MO-2ch}) by the $N/D$ technique \cite{Chew:1960iv}, as was done in Refs.~\cite{Danilkin:2020pak,Deineka:2024mzt}.

In the present $IJ=02$ case, however, one can still work with the unsubtracted version of Eq.(\ref{eq:MO-2ch}) by restricting the analysis to the two dominant $\pi\pi/K\bar{K}$ channels. Since there are no data for the phase of $t_{12}$, one needs to use the PDG information \cite{ParticleDataGroup:2024cfk} on the relevant resonances and their branching ratios to $\pi\pi$ and $K\bar{K}$. Similarly to the $\pi\pi/K\bar{K}$ analysis in the $IJ=00$ channel, the D-wave is controlled by two resonances, $f_2(1270)$ and $f_2'(1525)$, which, according to the PDG, couple to both channels but each decays dominantly to one of them. Therefore, the constraint $\sum \delta_a \to 2\pi$ is naturally realized in the $K$-matrix model discussed in the next subsection.

\begin{figure*}[t]
\centering
\includegraphics[width = 0.33\textwidth]{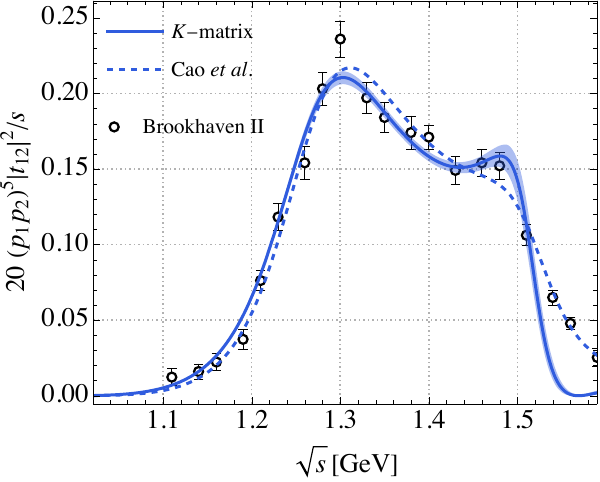}
\includegraphics[width = 0.33\textwidth]{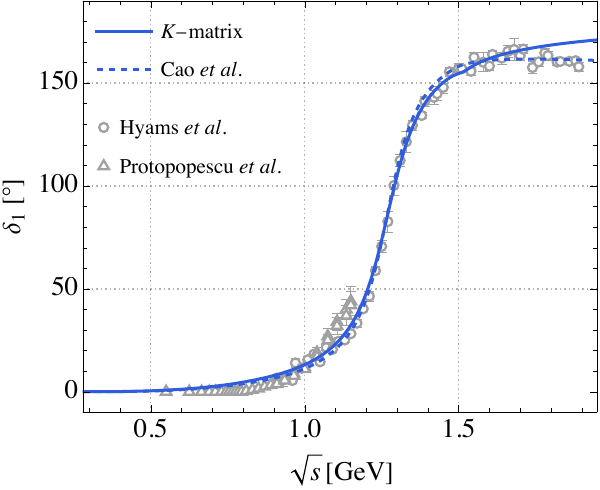}
\includegraphics[width = 0.33\textwidth]{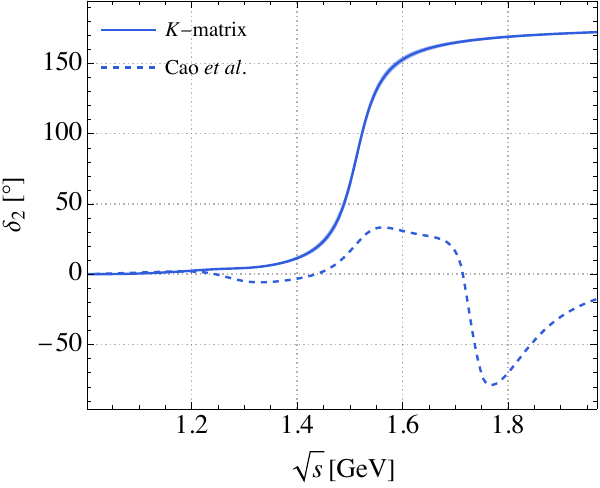}
\caption{Comparison of the minimal two-pole $K$-matrix fit (solid curves; bands show the $1\sigma$ fit uncertainty from the Brookhaven II fit) 
with the input used in Cao \emph{et al.}~\cite{Cao:2025dkv} (dashed curves) for the quantities entering 
the MO analysis: $|t_{12}|$, $\delta_1$, and $\delta_2$. The analyses of Ref.~\cite{TarrusCastella:2021pld} employ input very similar to that of Ref.~\cite{Cao:2025dkv} and are therefore not shown separately. Brookhaven II data~\cite{Longacre:1986fh} are shown for the quantity $20\,(p_1\,p_2)^{5}\,|t_{12}|^2/s$. The experimental data for $\delta_1$ are taken from \cite{Hyams:1973zf,Protopopescu:1973sh}.}
\label{Fig:K-matrix}
\end{figure*}

\begin{table*}[t]
  \centering
  \setlength{\tabcolsep}{5pt}
  \begin{tabular}{lccccccc}
    \hline\hline
    & \multicolumn{3}{c}{$K$-matrix parameters} 
    & \multicolumn{3}{c}{Pole position and couplings} 
    & \multicolumn{1}{c}{From pole} \\
    \cline{2-8}
      & $m_R$ [MeV] & $\Gamma_R$ [MeV] 
      & $g_{R,\pi\pi},\,g_{R,K\bar K}$, GeV$^{-1}$
      & $M_p$ [MeV] 
      & $\Gamma_p$ [MeV] 
      & $|g_{p,\pi\pi}|,\,|g_{p,K\bar K}|$, GeV$^{-1}$
      & $\mathcal{B}_{\pi\pi},\,\mathcal{B}_{K\bar K}$, $\%$ \\
    \hline
    $f_2(1270)$   & $1275.4(8)$ & $185.8(2.8)$ & $1.25(1)$, $0.77(1)$ & $1267.1(0.8)$ & $185.7(2.5)$ & $1.26(1)$, $0.81(1)$ &  $96.3(1),\,4.2(1)$ \\
    $f'_2(1525)$  & $1517.3(2.4)$ & $72(7)$ & $0.09(1),\,-1.14(5)$ & $1514.6(2.5)$ & $71.5(6.4)$ & $0.11(1),\,1.14(5)$ & $3.1(2),\,96.8(2)$ \\
    \hline\hline
  \end{tabular}
\caption{Resonance parameters from the minimal two-pole $K$-matrix fit to Brookhaven II data \cite{Longacre:1986fh}. For each resonance we list the $K$-matrix mass $m_R$ and width $\Gamma_R$ (both fixed to the PDG values \cite{ParticleDataGroup:2024cfk}), the fitted couplings $g_{R,a}$ (see text for more details), as well as the extracted pole position $\sqrt{s_p}=M_p-i\Gamma_p/2$, the moduli of the pole couplings $|g_{p,a}|$, and the branching fractions $\mathcal{B}_{a}$ obtained from these pole residues according to Eq.~(\ref{eq:Br_a}). The correlations among the fitted and derived (via Eq.~(\ref{eq:GammaR})) couplings are summarized in Table~\ref{tab:corr_matrix}.} 
\label{tab:K-matrix}
\end{table*}

\begin{table}[ht]
\centering
\begin{tabular}{l|cccc}
\hline \hline
 & $g_{f_2,\pi\pi}$ & $g_{f_2,KK}$ & $g_{f'_2,\pi\pi}$ & $g_{f'_2,KK}$ \\
\hline
$g_{f_2,\pi\pi}$ & $1.00$ & & & \\
$g_{f_2,KK}$ & $+0.29$ & $1.00$ & & \\
$g_{f'_2,\pi\pi}$ & $-0.15$ & $-0.52$ & $1.00$ & \\
$g_{f'_2,KK}$ & $+0.01$ & $-0.15$ & $+0.35$ & $1.00$ \\
\hline \hline
\end{tabular}
\caption{Correlation matrix of the resonance couplings derived from the bootstrap analysis. }
\label{tab:corr_matrix}
\end{table}

\subsection{K-matrix model}
\label{subsec:K-matrix}
To obtain an explicit scattering $\boldsymbol{t}$ matrix that satisfies coupled-channel unitarity and reproduces the available data, we employ a $K$-matrix approach. We write
\begin{equation}\label{eq:K-matrix}
\boldsymbol{t}^{-1}(s) = \boldsymbol{K}^{-1}(s)  -\, i\, \boldsymbol{\rho}(s)\,,
\end{equation}
which automatically satisfies the unitarity relation of Eq.~\eqref{eq:Imtinv}, provided that on
the right-hand cut the $\boldsymbol{K}$-matrix is real. Time-reversal invariance implies that the scattering amplitude is symmetric, so the $K$-matrix must also be symmetric, $K_{ab}(s)=K_{ba}(s)$. Typically, $K(s)$ is parametrized as a sum of pole terms describing resonances, supplemented by a smooth background contribution. For the isoscalar D-wave $\pi\pi/K\bar K$ system we include two poles, associated with the $f_2(1270)$ (labeled $R=1$) and the $f'_2(1525)$ ($R=2$),
\begin{equation}\label{eq:Kparam}
K_{ab}(s) = \sum_{R=1}^{2} \frac{\tilde{g}_{R,a}(s)\,\tilde{g}_{R,b}(s)}{m_R^2 - s}\,.
\end{equation}
For $J>0$ it is customary to include Blatt--Weisskopf barrier factors. These preserve the correct threshold behavior while suppressing unphysical growth at high energies, leaving the asymptotic behavior of the scattering amplitudes similar to the S-wave case. We include the D-wave Blatt--Weisskopf barrier factors in the $K$-matrix couplings $g_{R,a}(s)$ to shorten the notation. Specifically, we take
\[
\tilde{g}_{R,a}(s) = g_{R,a} \,
\frac{F_2(p_a(s)\,R_{int})}{F_2(p_a(m_R^2)\,R_{int})} \, ,
\]
where $g_{R,a}$ are fit constants and
\[
F_2(x) = \sqrt{\frac{1}{x^4+3 x^2+9}}\,.
\]
The interaction
radius is fixed to $R_{\text{int}} = 3.62~\text{GeV}^{-1}$ for both resonances, in line with analyses of D-wave resonances by Belle \cite{Belle:2007ebm, Belle:2013eck}.

The constraints given in Eq.~(\ref{eq:constraints}) imply, in our case, that $\delta_{12}(\infty) = 2\pi$ and $\eta(\infty) = 1$. In a $K$-matrix with two poles located above both thresholds and coupled to both channels, this asymptotic behavior can be implemented in a natural way. In the present work, we adopt the simplest realization with the standard two-body phase-space matrix and a two-pole $K$-matrix without background terms (as indicated in Eqs.~(\ref{eq:K-matrix}) and~(\ref{eq:Kparam})). This leads to $\delta_{12}(s \to \infty) \simeq 2\pi - c/s$ with $c>0$. An advantage of this construction is that the $K$-matrix parameters and the corresponding pole parameters turn out to be numerically close to each other. For one resonance this construction corresponds to a Flatt\'e form \cite{Flatte:1976xu}. In principle, one may also use the Chew--Mandelstam phase-space function \cite{Chew:1960iv, Wilson:2014cna}, which has a logarithmically growing real part above threshold. In that case polynomial background terms can be included in $K(s)$ without changing the required asymptotic limit. We do not consider this generalization in the present analysis. In both cases, the way in which the total asymptotic phase $2\pi$ is distributed between the individual channel phases, $\delta_1(\infty)$ and $\delta_2(\infty)$, is determined by the fitted reduced couplings $\tilde g_{R,a}$, which ultimately fix the corresponding resonance branching ratios.

\section{Numerical solution of the MO equations}
\label{sec:num}
\begin{figure*}[t]
\centering
\includegraphics[width =0.245\textwidth]{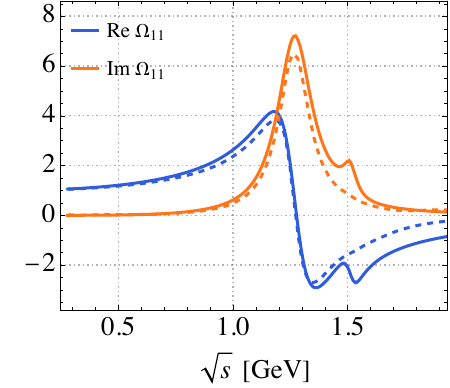}
\includegraphics[width =0.245\textwidth]{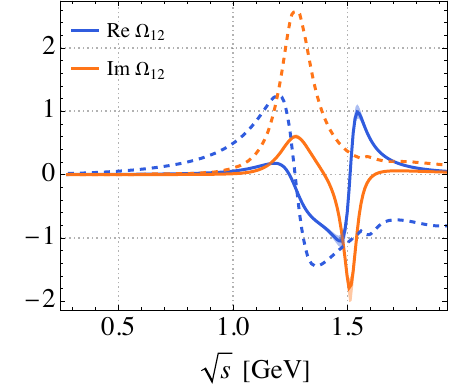}
\includegraphics[width =0.245\textwidth]{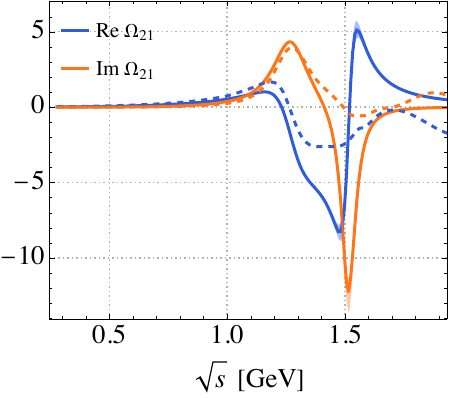}
\includegraphics[width =0.245\textwidth]{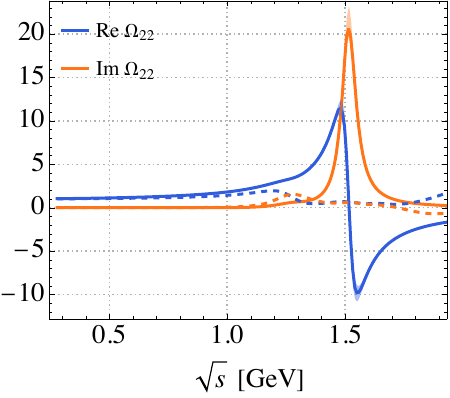}
\caption{Real and imaginary parts of the D-wave isoscalar $\pi\pi/K\bar{K}$ Omn\`es matrix elements $\Omega_{ab}(s)$. Solid curves show the result obtained from the two-pole $K$-matrix model used in this work, while dashed curves correspond to the Omn\`es matrix of Cao \emph{et al.}~\cite{Cao:2025dkv}. The shaded bands indicate the $1\sigma$ uncertainty propagated from the $K$-matrix input.
}
\label{Fig:Omnes}
\end{figure*}

To keep the Breit-Wigner parameters quoted by the PDG and the pole parameters as close as possible, we adopt a model with the standard two-body phase space and no background terms in the $K$-matrix. In this setup, residual contributions from subleading inelastic channels are absorbed by slightly adjusting the branching fractions. We fix the $K$-matrix masses $m_R$ to their PDG values \cite{ParticleDataGroup:2024cfk} (see Table~\ref{tab:K-matrix}) and determine the two-channel $K$-matrix couplings by imposing the constraint
\begin{equation}
\label{eq:GammaR}
    \Gamma_R=\sum_{a}\Gamma_a,\quad \Gamma_a=\frac{g_{R,a}^2}{m_R}\,\rho_{a}(m_R^2)\,,
\end{equation}
where $\Gamma_R$ is fixed to the PDG widths \cite{ParticleDataGroup:2024cfk} and $\rho_a(s)$ is the phase-space factor defined in Eq.~\eqref{eq:sigma}. This constraint leaves us with two unknown couplings (one per resonance) and the relative sign of $g_{R,1}\,g_{R,2}$ between $R_1$ and $R_2$. The latter is determined through the interference and is very well constrained by the Brookhaven II data on $|t_{12}|$ \cite{Longacre:1986fh}. With this minimal model we
obtain a good description of the Brookhaven II data up to $\sqrt{s}\simeq 1.5~\text{GeV}$, with $\chi^2/N_{\text{dof}} = 1.3$. In Table~\ref{tab:K-matrix} we also show pole parameters of the $t$-matrix
\begin{equation}
  t_{ab}(s) \;\sim\; \frac{g_{p,a}\,g_{p,b}}{s_p - s}\,,\quad \sqrt{s_p} \equiv M_p - \tfrac{i}{2}\,\Gamma_p\,,
  \label{eq:t-pole}
\end{equation}
where $g_{p,a}$ denotes the residue (pole coupling) to channel $a$; the complex-conjugate pole lies in the upper half-plane. Following the PDG prescription~\cite{ParticleDataGroup:2024cfk}, we compute the branching ratio of the resonance into channel $a$ from\footnote{We note that, within this prescription, the sum of the branching fractions is not constrained to be exactly 100\%.}
\begin{equation}
\label{eq:Br_a}
    \mathcal{B}_{a} =
\frac{1}{\Gamma_p}\left(\frac{|g_{p,a}|^2}{M_p}\rho_a(M_p^2)\right)\,.
\end{equation}
Since in our case both narrow resonances lie well above the $\pi\pi$ and $K\bar K$ thresholds, the poles closest to the physical region are located on Riemann sheet~III, characterized by $\operatorname{Im} p_1 < 0$, $\operatorname{Im} p_2 < 0$. Given the limitations of a two-channel description, the resulting branching fractions are reasonably close to the PDG estimates~\cite{ParticleDataGroup:2024cfk}. The pole mass of the $f_2(1270)$ lies slightly below the PDG Breit-Wigner mass, but well within the quoted PDG pole range \cite{ParticleDataGroup:2024cfk} (see also \cite{Pelaez:2025jrn,Pelaez:2024uav} for more recent developments). For comparison, in Fig.~\ref{Fig:K-matrix} we also show the input used in \cite{TarrusCastella:2021pld,Cao:2025dkv}. For $|t_{12}|$ and $\delta_{12}$ these works employ the CFD parametrizations of Refs.~\cite{Pelaez:2018qny,Pelaez:2020gnd}, for which the off-diagonal $D$-wave amplitude is given by
\footnote{Our normalization for the off-diagonal matrix element matches that of Refs.~\cite{Pelaez:2018qny,Pelaez:2020gnd}, i.e.\
$t_{12}(s) = g_2^0(s)$.}
\begin{equation}
t^{\mathrm{CFD}}_{12}(s) = 
\frac{C\left(
e^{i\phi_1} \mathrm{BW}_1(s) + \beta e^{i\phi_2} \mathrm{BW}_2(s) + \gamma e^{i\phi_3} \mathrm{BW}_3(s)
\right)}{
2\sqrt{1 + r_2^2 \hat{q}_\pi^4(s)} \, \sqrt{1 + r_2^2 \hat{q}_K^4(s)}
}\,,
\end{equation}
where $\mathrm{BW}_i(s)$ are Breit-Wigner-like shapes with
energy-dependent widths (see Refs.~\cite{Pelaez:2018qny,Pelaez:2020gnd} for details). As argued in the introduction, this model does not enforce coupled-channel unitarity. The CFD step adds analyticity/crossing via p.w. dispersion constraints, improving consistency, but it still does not turn the model into a unitary coupled-channel fit. The resulting phase $\delta_{12}(s)$ differs substantially from the unitary two-channel solution obtained from our $K$-matrix fit. To illustrate this difference, in Fig.~\ref{Fig:K-matrix} we show $\delta_2=\delta_{12}-\delta_1$, where the input for $\delta_1$ in \cite{Cao:2025dkv} is taken from \cite{Bydzovsky:2016vdx}. The latter is consistent with our $K$-matrix result up to $1.6~\text{GeV}$. The markedly different behavior of $\delta_2(s)$ translates into noticeable changes of the Omn\`es matrix, in particular in the elements involving the $K\bar K$ channel. In Fig.~\ref{Fig:Omnes} we compare the solution of the MO equation based on the $K$-matrix fit with the results from Cao \emph{et al.}~\cite{Cao:2025dkv}.

\section{\texorpdfstring{Application to $J/\psi\to\pi^{0}\pi^{0}\gamma,\,K_{S}K_{S}\gamma$}{Application to J/psi radiative decays}}\label{sec:Jpsi-application}

In this section we apply the coupled-channel Omn\`es matrix constructed above to the radiative decays $J/\psi\to\pi^{0}\pi^{0}\gamma$ and $J/\psi\to K_{S}K_{S}\gamma$. Both processes are dominated, in the D-wave, by the $f_{2}(1270)$ and $f_{2}'(1525)$ tensor-meson resonances and therefore constitute an ideal testing ground for the two-channel Omn\`es input. Our goal is to build a dispersive representation of the electric-dipole (E1), $J=2$ amplitudes and fit them directly to the BESIII data \cite{BESIII:2015rug,BESIII:2018ubj}. To avoid using different overall normalization constants for $J/\psi\to\pi^{0}\pi^{0}\gamma$ and $J/\psi\to K_{S}K_{S}\gamma$, we normalize the BESIII spectra as explained below.

\subsection{Experimental spectra}
In Refs.~\cite{BESIII:2015rug,BESIII:2018ubj}, the branching fraction is determined according to
\begin{equation}\label{eq:Branching}
\mathcal{B}_f
= \frac{I_{\mathrm{tot}}-N_{\mathrm{bkg}}}{\epsilon _{\gamma } N_{J/\psi}}\,,
\end{equation}
where $f$ is either $J/\psi\to\pi^{0}\pi^{0}\gamma$ or $J/\psi\to K_{S}K_{S}\gamma$. In Eq.~(\ref{eq:Branching}) $I_{\mathrm{tot}}$ is the number of acceptance- and efficiency-corrected events, $N_{\mathrm{bkg}}$ is the number of remaining background events\footnote{These backgrounds either survived the $\pm 50$ MeV veto around the $\omega$ mass in the $\gamma\pi^0$ invariant mass for $J/\psi\to\pi^{0}\pi^{0}\gamma$, or correspond to components that are not subtracted in the fit via negative-weight Monte Carlo samples. Their overall contribution is very small: about 2.3\% for $\pi\pi$ and 1.14\% for $K_S K_S$.}, $N_{J/\psi}$ is the number of produced $J/\psi$ mesons, and $\epsilon_{\gamma}=0.9993$ is related to the photon-energy extrapolation. For each final state, BESIII also reports the acceptance-corrected event yield $I_{\mathrm{tot},k}$ in each mass bin $k$ of width $\Delta M$. Since the bin-by-bin background distribution $N_{\mathrm{bkg},k}$ is not given in \cite{BESIII:2015rug,BESIII:2018ubj}, we assume that its shape is proportional to the signal (any distortion from the neglected bin dependence of the small residual backgrounds is expected to be at the percent level or below, well within the experimental uncertainties). Considering the unit-area (normalized) spectrum, which cancels global normalization factors, we obtain
\begin{equation}
\frac{1}{\mathcal{B}_{f}}
\left[\frac{d\mathcal{B}_{f}}{d\sqrt{s}}\right]_{k}=\frac{1}{\Gamma_{f}}
\left[\frac{d\Gamma_{f}}{d\sqrt{s}}\right]_{k}
\simeq 
\frac{I_{\mathrm{tot},k}}
{\Delta M\,I_{\mathrm{tot}}}\,,
\end{equation}
where $\Gamma_{f}$ denotes the partial decay width into the final state $f$. In Refs.~\cite{BESIII:2015rug,BESIII:2018ubj}, BESIII also provides mass-independent amplitude analyses for both final states. They decompose the spectra into contributions from total meson-meson angular momentum $J=0,2$, and from photon multipoles,
\begin{equation}
\frac{d\Gamma_{f}}{d\sqrt{s}}
=
\frac{d\Gamma^{(0),\mathrm{E1}}_{f}}{d\sqrt{s}}
+
\sum_{X=\mathrm{E1,M2,E3}}
\frac{d\Gamma^{(2),X}_{f}}{d\sqrt{s}}\,,
\end{equation}
or, in terms of event yields,
\begin{equation}
    I_{\mathrm{tot},k}
= I^{(0),\mathrm{E1}}_{k} +\sum_{X=\mathrm{E1,M2,E3}} I^{(2),X}_{k}\,.
\end{equation}
In this work we focus exclusively on the dominant $J=2$, E1
term. Therefore, the central observable entering our fits in the
next subsection is
\begin{equation}
  \frac{1}{\Gamma_f}\frac{d\Gamma_f^{(2),\text{E1}}}{d\sqrt{s}}.
  \label{eq:obs}
\end{equation}

\subsection{\texorpdfstring{Dispersive model for the $J=2$, E1 amplitudes}{Dispersive model for the J=2, E1 amplitudes}}

The angular integrations over the three-body phase space can be done analytically using the orthogonality of Wigner $d$-functions. Denoting $q\equiv M_{J/\psi}$, the differential widths for the $J=2$, E1 partial wave read
\begin{align}
\frac{d\Gamma^{(2),\mathrm{E1}}_{J/\psi\to\pi^{0}\pi^{0}\gamma}}{d\sqrt{s}}
&=
\frac{(q^{2}-s)\,\sqrt{s-4m_{\pi}^{2}}}{384\pi^{3}q^{3}}
\,
5\,\left|h^{(2),\mathrm{E1}}_{\pi^{0}\pi^{0}}(s)\right|^{2},
\\[4pt]
\frac{d\Gamma^{(2),\mathrm{E1}}_{J/\psi\to K_{S}K_{S}\gamma}}{d\sqrt{s}}
&=
\frac{(q^{2}-s)\,\sqrt{s-4m_{K}^{2}}}{384\pi^{3}q^{3}}
\,
5\,\left|k^{(2),\mathrm{E1}}_{K^{0}\bar K^{0}}(s)\right|^{2},
\end{align}
where, for $K_S K_S$, we neglect tiny CP violation and use that, for even $J$, the $K_S K_S$ and $K^0 \bar{K}^0$ partial waves coincide. The usual symmetry factor for the identical $\pi^0\pi^0$ and $K_S K_S$ final state is taken into account in both expressions. The physical amplitudes are expressed in terms of isoscalar amplitudes as
\begin{equation}
  h_{\pi^0\pi^0}(s) = -\frac{1}{\sqrt{3}}\,h_{0}(s)\,,
  \quad
  k_{K^0\bar K^0}(s) = -\frac{1}{2}\,k_{0}(s)\,,
\end{equation}
where the additional $1/\sqrt{2}$ factor in the $K\bar K$ channel ensures a consistent unitarity normalization between partial waves with identical and non-identical particles in the isospin basis (see, e.g., Ref.~\cite{Garcia-Martin:2010kyn} for the $S$-wave case). As in the scattering case, we factor out the trivial kinematic factors associated with the $D$-wave threshold behavior and the soft-photon limit,
\begin{equation}
\begin{pmatrix}
h^{(2),\mathrm{E1}}_{0}(s)\\
k^{(2),\mathrm{E1}}_{0}(s)
\end{pmatrix}
\equiv (s-q^2)\,
\begin{pmatrix}
p_1^2(s) & 0 \\
0 & p_2^2(s)
\end{pmatrix}
\begin{pmatrix}
\hat{h}(s) \\
\hat{k}(s)
\end{pmatrix}\,,
\end{equation}
where $\hat{h}(s)$, $\hat{k}(s)$ are free of kinematic constraints. The isoscalar production vector then satisfies the following two-channel unitarity relation
\begin{equation}
    \operatorname{Im}
\begin{pmatrix}
\hat{h}(s) \\
\hat{k}(s)
\end{pmatrix}
=
\boldsymbol{t}(s)\, \boldsymbol{\rho}(s)\,
\begin{pmatrix}
\hat{h}(s) \\
\hat{k}(s)
\end{pmatrix}^{*}\,,
\end{equation}
with $\boldsymbol{t}(s)$ and $\boldsymbol{\rho}(s)$ as defined in Sec.~\ref{sec:Formalism}. We can then write a Muskhelishvili-Omn\`es representation based on the Omn\`es matrix $\boldsymbol{\Omega}(s)$, which satisfies the unitarity relation given in Eq.~(\ref{eq:MO-2ch}). Writing a dispersion relation for $\boldsymbol{\Omega}^{-1}(s)\left(
\hat{h}(s)\,,
\hat{k}(s)
\right)^T$,
the unsubtracted standard MO representation takes the following form\footnote{Note that for $s>s_{\mathrm{th}1}$ the relation $\text{Im}\left[\,\boldsymbol{\Omega}^{-1}(s)\left(
\hat{h}(s)\,,
\hat{k}(s)
\right)^T\right]=0$ is satisfied only if in $\boldsymbol{t}(s)$ and $\hat{h}(s),\,\hat{k}(s)$ the threshold factors have been pulled out in a consistent way, i.e. they are free of kinematic constraints.}
\begin{equation}
    \begin{pmatrix}
\hat{h}(s) \\
\hat{k}(s)
\end{pmatrix}=
\boldsymbol{\Omega}(s)\,\left[\int_{L}\frac{ds'}{\pi}\frac{\boldsymbol{\Omega}^{-1}(s')}{s'-s}\,\text{Im}\begin{pmatrix}
\hat{h}^{L}(s') \\
\hat{k}^{L}(s')
\end{pmatrix}\right]\,,
\end{equation}
where $L$ denotes the left-hand cut. Since the left-hand contribution in the physical region typically is a slowly changing function, we approximate it by the subtraction polynomial, which, in terms of the physical D-wave E1 amplitudes, leads to
\begin{equation}\label{eq:Jpsi-MO}
 \begin{pmatrix}
h^{(2),\mathrm{E1}}_{0}(s)\\
k^{(2),\mathrm{E1}}_{0}(s)
\end{pmatrix}
  \approx (s - q^2)\,
    \begin{pmatrix}
      p_1^2(s) & 0 \\
      0        & p_2^2(s)
    \end{pmatrix}
    \boldsymbol{\Omega}(s)
    \begin{pmatrix}
      a + b\,s \\[0.2em]
      c + d\,s
    \end{pmatrix}.
\end{equation}
Thus all non-trivial energy dependence induced by strong $\pi\pi/K\bar K$ is encoded in the Omn\`es matrix $\boldsymbol{\Omega}(s)$, while the four real parameters $(a,b,c,d)$ describe short-distance physics and left-hand-cut effects. As will be shown below, the data do not indicate the need for higher-order terms.

\subsection{Results of the fit}
\label{subsec:Jpsi-fit}

We now discuss the outcome of the simultaneous fit of the dispersive model in Eq.~(\ref{eq:Jpsi-MO}) to the normalized $J=2$, E1 spectra of $J/\psi \to \pi^0\pi^0\gamma$ and $J/\psi \to K_S K_S\gamma$ in the range\footnote{We choose $\sqrt{s}\le 1.6~\mathrm{GeV}$ to stay in the region where a two-channel description is expected to be most reliable; extending the fit window (e.g., to $\sim 1.75~\mathrm{GeV}$) does not change the conclusions.} $\sqrt{s} \le 1.6~\text{GeV}$. We note that the four real parameters $(a,b,c,d)$ of the production polynomial are taken to be common to both final states, while the Omn\`es matrix is fixed as described in Sec.~\ref{sec:MO}. As input we also fixed $\Gamma_{J/\psi} =92.6$ keV, $\mathcal{B}_{J/\psi \to \pi^0\pi^0\gamma}=1.15038\times 10^{-3}$, and $\mathcal{B}_{J/\psi \to K_S K_S\gamma}=8.10278\times 10^{-4}$ \cite{ParticleDataGroup:2024cfk}.

\begin{figure}[t]
\includegraphics[width =0.475\textwidth]{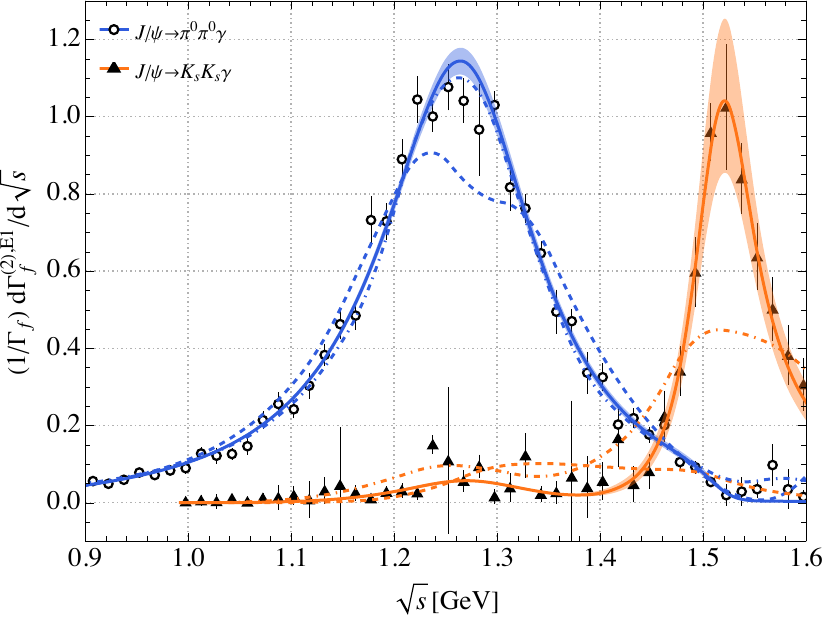}
\caption{Combined fit to the normalized $J=2$, E1 partial-wave spectra in the decays $J/\psi \to \pi^0\pi^0\gamma$ and $J/\psi \to K_S K_S\gamma$. The data points with error bars show the BESIII mass-independent analysis \cite{BESIII:2015rug,BESIII:2018ubj}, while the solid curves represent the coupled-channel Muskhelishvili-Omn\`es fit based on Eq.~(\ref{eq:Jpsi-MO}). Shaded bands show the $1\sigma$ uncertainty from the fit and the propagated $K$-matrix input uncertainties.
For comparison, we also show the results using Omn\`es matrices of Tarr{\'u}s Castell{\`a} \emph{et al.}~\cite{TarrusCastella:2021pld} (dashed) and Cao \emph{et al.}~\cite{Cao:2025dkv} (dot-dashed).
} 
\label{Fig:Jpsi}
\end{figure}

The resulting curves are shown in Fig.~\ref{Fig:Jpsi} and the numerical values of the fitted parameters are summarized in Tab.~\ref{tab:Jpsi_params}. One observes that the coupled-channel MO representation reproduces simultaneously the shape and normalization of the $f_2(1270)$ peak in the $\pi^0\pi^0$ channel and of the $f_2'(1525)$ peak in the $K_S K_S$ channel with $\chi^2/N_{\text{dof}}=1.1$. In particular, the interference pattern in the region between the two resonances is well described, indicating that the energy dependence generated by the $\pi\pi/K\bar{K}$ final-state interactions is correctly captured by our Omn\`es matrix. We have also tested the subleading M2 and E3 multipoles, using a slightly modified MO representation where the soft-photon factor $(s - q^2)$ is generalized to $(s - q^2)^L$ with $L=2,3$ for M2 and E3, respectively. The resulting fits are of similarly good quality. Since the M2 and E3 data have larger uncertainties and do not add qualitatively new information, we do not discuss them further.

\begin{table}[ht]
\centering
\begin{tabular}{c c c c c}
\hline\hline
 $a/\sqrt{5}$ & $b/\sqrt{5}$ & $c/\sqrt{5}$ & $d/\sqrt{5}$ & $\chi^{2}/N_{\text{dof}}$ \\
\hline
 $-2.11(4)$ 
& $0.72(5)$
& $-1.61(32)$
& $0.39(14)$
& $1.1$\\
\hline\hline
\end{tabular}
\caption{Parameters of the production polynomial defined in Eq.~(\ref{eq:Jpsi-MO}) (in units of $10^{-3}~\text{GeV}^{-4}$ for $a,c$ and $10^{-3}~\text{GeV}^{-6}$ for $b,d$) obtained from the combined fit to the normalized $J=2$, E1 spectra of $J/\psi \to \pi^0\pi^0\gamma$ and $J/\psi \to K_S K_S\gamma$ in the range $\sqrt{s} \le 1.6~\text{GeV}$.}
\label{tab:Jpsi_params}
\end{table}

To quantify the impact of the Omn\`es input, we have repeated the analysis using the coupled-channel Omn\`es matrices of Refs.~\cite{TarrusCastella:2021pld} and~\cite{Cao:2025dkv}. The corresponding values of $\chi^2/N_{\rm data}$ are collected in Table~\ref{tab:comp}, where we show separately the contributions from the $\pi\pi$ and $K\bar{K}$ channels and from the combined data set, for three fit strategies: a simultaneous $\pi\pi+K\bar{K}$ fit, a $\pi\pi$-only fit, and a $K\bar{K}$-only fit. In all possible fits our Omn\`es matrix achieves significantly better $\chi^2$ values, in particular for the $K\bar{K}$ spectrum, reflecting physically unreasonable input for the $K\bar{K}$ phase shift used in \cite{TarrusCastella:2021pld,Cao:2025dkv} (see right panel of Fig.~\ref{Fig:K-matrix}). As for the $\pi\pi$ spectrum, the fits using $\delta_1(\infty)\to \pi$ \cite{Cao:2025dkv} are clearly better than the one with $\delta_1(\infty)\to 2\pi$ \cite{TarrusCastella:2021pld}.

\begin{table}[t]
\centering
  \begin{tabular}{llllc}
    \hline\hline
    Fit type & $\chi^2/N_{\rm data}$ 
             & Ref.~\cite{TarrusCastella:2021pld} 
             & Ref.~\cite{Cao:2025dkv}
             & This work \\[0.2em]
    \hline
    \multirow{3}{*}{$\pi\pi$+$K\bar K$}
      & $\pi\pi$               & 2.7 & 1.3 & 1.1 \\
      & $K\bar K$              & 12.4 & 4.5 & 1.0 \\
      & total (fit)                      & 5.8 &  2.3 & 1.1\\
      \hline 
    \multirow{3}{*}{$\pi\pi$-only}
      & $\pi\pi$ (fit)              & 2.7 & 1.1 & 1.1 \\
      & $K\bar K$ (post)            & 12.8 & 136 & 4.1 \\
      & total                       & 5.9 & 44 & 2.1 \\ \hline 
    \multirow{3}{*}{$K\bar K$-only}
      & $K\bar K$ (fit)             & 2.6 & 2.9 & 0.9 \\
      & $\pi\pi$  (post)             & $\sim 10^6$ & 44 & 3.2 \\
      & total                        & $\sim 10^6$ & 31 & 2.5 \\
    \hline\hline
  \end{tabular}
  \caption{Quality of the description of the $J=2$, E1 $J/\psi\to\pi^0\pi^0\gamma$ and $J/\psi\to K_S K_S\gamma$ spectra using different Omn\`es matrices: Tarr{\'u}s Castell{\`a} \emph{et al.}~\cite{TarrusCastella:2021pld}, Cao \emph{et al.}~\cite{Cao:2025dkv}, and this work. We quote $\chi^2/N_{\rm data}$ separately for the $\pi\pi$ and $K\bar K$ channels and for the combined data set ($N_{\rm tot}=N_{\pi\pi}+N_{K\bar K}$). Superscripts ``fit'' and ``post'' indicate, respectively, channels included in the fit and channels predicted without refitting.}
\label{tab:comp}
\end{table}

\section{Summary}

In this work we have constructed a D-wave isoscalar $\pi\pi/K\bar K$ coupled-channel Omn\`es matrix based on a unitary two-channel $K$-matrix model with two poles associated with the $f_2(1270)$ and $f'_2(1525)$ resonances. The model uses PDG masses and widths as input, and the couplings are fitted to $\pi\pi \to K\bar K$ data. The resulting unitary scattering amplitude is then used as input to the homogeneous Muskhelishvili-Omn\`es problem. This construction addresses several limitations of earlier approaches based on Breit-Wigner sums, in particular the lack of exact coupled-channel unitarity and the implausible behavior of the implied $K\bar K$ phase shift.

As an application, we have implemented this Omn\`es matrix in a dispersive representation of the $J = 2$ electric-dipole amplitudes for the radiative decays $J/\psi \to \pi^0\pi^0\gamma$ and $J/\psi \to K_S K_S \gamma$, and fitted the resulting expressions to the BESIII mass-independent analyses of the corresponding spectra. Our Omn\`es input allows for a simultaneous description of both channels with $\chi^2/N_{\text{dof}} \simeq 1.1$, correctly reproducing the $f_2(1270)$ and $f'_2(1525)$ peaks and the interference pattern in between. A systematic comparison with the coupled-channel Omn\`es matrices of Refs.~\cite{TarrusCastella:2021pld,Cao:2025dkv}, summarized in Table~\ref{tab:comp}, shows that our solution not only yields the best overall fit quality, but also exhibits clearly superior predictive power when only one of the two channels is used as input.

These results strongly support the consistency of our coupled-channel description of D-wave $\pi\pi/K\bar K$ final-state interactions. The Omn\`es matrix constructed here can serve as a controlled dispersive input for future studies of tensor form factors and resonance properties, and can be straightforwardly applied to other processes sensitive to the $f_2(1270)/f'_2(1525)$ system.

\section*{Acknowledgements}
We are grateful to Bachir Moussallam for providing the MO code for the single-channel $\rho$-meson case, which clarified key numerical aspects of Ref.~\cite{Moussallam:1999aq}.
This work was supported by the Deutsche Forschungsgemeinschaft (DFG, German Research Foundation) within the Research Unit [Photon-photon interactions in the Standard Model and beyond, Projektnummer 458854507 - FOR 5327]. The work of EP is supported by the Spanish Government (Agencia Estatal de Investigaci\'on MCIN/AEI/10.13039/501100011033)
Grants No. PID2020--114473GB-I00, No. PID2023-146220NB-I00, and CEX2023-001292-S
(Agencia Estatal de Investigaci\'on MCIU/AEI (Spain) under grant IFIC Centro de Excelencia
Severo Ochoa), by the Generalitat Valenciana (Spain) through the plan
GenT program CIDEGENT/2021/037 and by the U.S. National Science
Foundation under grant PHY-2310149. 

\bibliographystyle{apsrevM}
\bibliography{1}

\end{document}